\documentclass{article}
\usepackage{amssymb}

\usepackage{graphicx}
\usepackage{amsmath}

\newtheorem{theorem}{Theorem}
\newtheorem{acknowledgement}[theorem]{Acknowledgement}

\input{tcilatex}

\begin{document}

\title{\textbf{Are spacetime horizons higher dimensional sources of energy fields? }%
\\
(\textbf{The black hole case).}}
\author{Manasse R. Mbonye \\
\\
\textit{Michigan Center for Theoretical Physics}\\
\textit{Physics Department, University of Michigan, Ann Arbor, MI 48109}}
\date{\today}
\maketitle

\begin{abstract}
We explore the possibility that spacetime horizons in 4D general relativity
can be treated as manifestations of higher dimensions that induce fields on
our 4D spacetime. In this paper we discuss the black hole event horizon, as
an example (we leave the cosmological case for future discussion). Starting
off from the field equations of gravity in 5D and some conditions on the
metric we construct a spacetime whose imbedding is a 4D generalization of
the Schwarzchild metric. The external region of the imbedded spacetime is
found to contain two distinct fields. We discuss the properties of the
fields and the potential implications. Taken as they are, the results
suggest that the collapse of matter to form a horizon may have non-local
consequences on the geometry of spacetime. In general, the use of
horizon-confined mass as a coordinate suggests three potential features of
our universe. The first is that the observed 4D spacetime curvature and
ordinary matter fields may be hybrid features of 5D originating from the
mixing of coordinates. Secondly, because the fifth coordinate induces
physical fields on the 4D hyperface, the global metric of the universe may
not be asymptotically flat. And finally, associating matter with an
independent dimension points towards a theory of nature that is scale
invariant.

\ 

\ 
\end{abstract}

\section{\protect\smallskip Introduction}

\smallskip In an effort to build a unified theory of nature to explain the
observed universe, physics has had to look up to higher dimensions. A
consequence of this effort has been the development, in recent years, of
higher dimensional models, notably the Superstring models$^{1}$ and M-Theory$%
^{2}$. In these models dimensions are usually assumed to be curled up into
length scales of order of Planck size, hence rendering them difficult to
observe\ at any energies lower than the Planck energy. This modern view of
higher dimensional compactification has its historical roots in \ Klein's
original interpretation$^{3}$ of (the stronger) Kaluza's \textit{cylinder
condition}$^{4}$\textit{\ (}which we discuss later\textit{).} Lately,
alternatives to compactification such as the Randall-Sundrum mechanism$^{5}$
have been proposed.

Dimensional Compactification is\textit{\ }certainly consistent with why we
don't sense the higher dimensions. Whether such a feature is also necessary
or whether nature may have other options is still academic. It is however
reasonable and may be prudent, at this time, to explore further alternative
possibilities. In this respect, one is reminded of the citizens of the
legendary 2D world in \textit{Flatland}$^{6}$ and their limitations in
perception. These citizens' understanding of concepts like angular momentum,
or the force on an electric charge moving in a magnetic field, requires them
to develop a higher dimensional theory involving cross-products. Clearly,
the `other' dimensions that form part of the manifold in which Flatland is
imbedded do not have to be compact. Flatlanders are just two dimensional
beings living in their two dimensional world with their observations in the
third dimension restricted by what they have learned to call a horizon. As a
result, Flatlanders would interpret effects of their interactions with the
higher dimensional world as originating from the horizon.

In this discussion we seek for possible manifestations and implications of
higher dimensions based on the assumption that such dimensions are not
necessarily compact. In our approach we represent matter as the fifth
coordinate, provided such matter is confined in a spacetime horizon. The
fifth coordinate is then given by the length $x^{4}=h$ associated with the
horizon size. The criterion employed in this work to treat a quantity as
coordinate representing a dimension, is the quantity's independence from the
other spacetime dimensions. In so far as a spacetime horizon (local or
global) signifies a boundary to our observable 4D universe, then the length $%
h$ associated with the horizon size meets the above criterion and its
treatment as a manifestation of a dimension can be justified on this basis.

\smallskip The possibility of representing matter as a coordinate has been
suggested before by Lessner$^{7}$ and by Wesson$^{8}$. It is however
difficult to justify the concept of a fifth dimension without the above
criterion, namely the independence of the associated coordinate. From the
onset it is clear, for example, that in our approach ordinary matter cannot
be justifiably treated as an independent coordinate. This is because the
worldline of any such matter (including particles) is associated with a 4D
spacetime measure and is therefore not independent of the spacetime
coordinates. Such an identification distinguishes our treatment, its results
and implications, from the previous treatments. As we point out in the next
section, ordinary matter (i.e. not trapped inside an event horizon) is seen
to be a hybrid, resulting from the mixing of spacetime coordinates with the
fifth coordinate.

We shall demonstrate that given a 5D spacetime geometry in which the fifth
coordinate $x^{4}$ is treated as a length $h$ corresponding to the horizon
size, then there is an imbedded 4D spacetime on which fields are induced. In
particular, taking the horizon size as that of a black hole mass $M$ so that 
$h=\frac{2GM}{c^{2}}$ we find that, in the case of the induced 4D geometry,
the external space outside the black hole is filled with two distinct
fluids. We discuss the properties of these fields and their implications.

\section{The 5D field equations}

\subsection{The fifth dimension}

\smallskip We begin by introducing the physical basis for the fifth
coordinate. One can build an intuitive sense of the fifth dimension by
drawing an analogy with the time dimension$^{9}$. In the Minkowiski
spacetime, the time dimension is associated with a length coordinate written
as $x^{0}=ict$, where $c$ is a universal constant (the upperbound speed for
propagation of physical information) . Because of the large value of this
constant, we do not usually sense time as a dimension except at high
velocities (comparable to $c$) where such relativistic phenomena as length
contraction and time dilation manifest themselves.

Analogously, in our present approach, one can associate the fifth dimension
with a coordinate 
\begin{equation}
x^{4}=2\sqrt{\varepsilon }\varkappa M=\sqrt{\varepsilon }h,  \tag{1}
\end{equation}
constructed from a mass $M$ confined in an event horizon of size $h$. Here,
the parameter $\varepsilon =\pm 1$ identifies the coordinate as space-like
or time-like and $\varkappa =\frac{G}{c^{2}}\approx \frac{l_{p}}{m_{p}}$ is
a universal constant with units of length per unit mass and gives the length
scale physically associated with maximum untrapped mass$.$ In MKS units $%
\varkappa =7.42\times 10^{-28}mkg^{-1}$. Because of the small value of $%
\varkappa $, the observable effects of the fifth dimension (analogous to
time dilation/length contraction) should be favored by high density fields.
One notes, however, that $\varkappa $ is independent of velocity. Thus, even
at ordinary densities (e.g. earth density) effects of higher dimensions
should be readily observable in our universe, as deviations from Minkowiski
spacetime, provided a large enough interval associated with the fields is
considered. With regard to the specific nature of the
(trapped-mass)-spacetime coordinate mixing effects, one expects a bending of
spacetime analogous to the special relativistic length contraction and a
`dilution' effect of the mass coordinate analogous to the time dilation.
Such effects would give rise to spacetime curvature and introduce matter
fields in 4D. The suggestive implication, then, is that ordinary 4D
spacetime curvature and ordinary matter/radiation are manifestations of
higher dimensions.

That curvature is imposed on the 4D spacetime is a known feature of general
relativity. The only addition our treatment suggests is that such imposition
results from the mixing of the fifth coordinate with the spacetime ones. On
the other hand, the observation that ordinary matter fields in 4D also
originate from such coordinate mixing is a broad concept to be a subject of
this initial and future discussions. We will also comment on the
cosmological and other implications of these effects later in the conclusion
of this discussion.

\subsection{The equations}

To proceed, we first a give a summary of the equations which set the
framework for the remaining discussion. These equations can be derived from
a five dimensional action of the form 
\begin{equation}
S=\frac{1}{16\pi \bar{G}}\int \sqrt{-g}\left( R+\ell \right) d^{5}x  \tag{2}
\end{equation}
by varying the former with respect to the 5D metric $g_{ab}$. Here $R$ and $%
\bar{G}$ are the 5D Ricci scalar and Newton's constant, respectively and $g$
is the determinant of the full 5D metric whose line element is 
\begin{equation}
dS^{2}=g_{ab}dx^{a}dx^{b},\;\left( a,b=0,..,4\right) .  \tag{3}
\end{equation}
Further, $\ell $ is a Lagrangian density which is invariant with respect to
arbitrary transformations of all the five coordinates $x^{a}$ (including the
mass coordinate $h=2\varkappa M$) and represents the fields that may result
(as pointed out above) from the mixing of these coordinates. In our
notation, lower case lettering is used for spacetime indices (Roman for 5D
and Greek in 4D) and from now on we stick to the geometrized units $8\pi
G=c=1$, unless otherwise stated.

As already mentioned above, our discussion is based on the assumption that
spacetime horizons can be treated as manifestations of higher dimensions.
With this assumption we investigate the possibility that such horizons are
generators \ of matter fields on the 4D spacetime.\textit{\ }One does expect
that the field equations derived from Eq. (2) should give a 5D\ geometry $%
g_{ab}$\ whose foliation yields a family of 4-hyp$\func{erf}$aces\ with
physically meaningful interpretation. In particular, we suppose in this
particular discussion that the metric induced on the 4D hyperface describes
the spacetime outside a \textit{static} source which is represented by the
horizon size. This puts constraints on the properties of such a metric
induced on the 4D manifold, and hence on the foliation character of the 5D
space. Thus we expect that the resulting 4D spacetime should:

\begin{itemize}
\item  (i) \ \ have spherical symmetry;

\item  (ii) \ reduce to the Schwarzschild solution$^{10}$ on a constant $h$
surface;

\item  (iii) be asymptotically flat (as $h\longrightarrow 0$).
\end{itemize}

In passing, one notes with regard to condition (iii), that since in a
cosmological sense $h$ never realistically goes to zero, and since its
expected effect is to induce fields on 4D spacetime, then the geometry of
the universe could never be globally flat. This suggests, independently,
that the 4D universe should, necessarily, be bathed in some vacuum, $\Lambda 
$, with an associated non-vanishing energy density $\rho _{\Lambda }$. This
issue will be revisited in a future discussion of the cosmological case.

Condition (i) demands that the geometry of the full 5D manifold should admit
a metric of the form 
\begin{equation}
dS^{2}=e^{\nu }dt^{2}-e^{\mu }dr^{2}-R^{2}d\Omega ^{2}+\varepsilon e^{\psi
}dh^{2},  \tag{4}
\end{equation}
where, condition (ii) implies that the metric coefficients can, at most, be
functions of $r$ and $h$ only.

Before proceeding, it is worthwhile to comment on this dependence of the
induced 4-metric $g_{\mu \nu }$\ on the fifth coordinate, $h$. In his
original work, Kaluza$^{4}$\ assumed that geometric and physical objects in
4D are independent of the fifth coordinate. This assumption is expressed in
the \textit{cylinder condition} which holds that all derivatives of the 4D
metric with respect to the fifth coordinate, must vanish, i.e. $\partial
_{\left( a=4\right) }\left[ g_{\mu \nu }\right] =0$. Later, Klein$^{3}$
introduced a weaker condition by letting the full 4-metric $g_{\mu \nu
}\left( x^{\mu },x^{4}\right) =g_{\mu \nu }^{0}\left( x^{\mu }\right)
X\left( x^{4}\right) $, be separable \ in the variables $x^{\mu }$ and a
compactified $x^{4}$. This view has been introduced in the modern string
theories. It holds that the higher dimensions are curled up into a length
scale of order of Planck size, hence rendering them difficult to observe\ at
any energies lower than the Planck energy. Throughout our treatment the 
\textit{cylinder condition} is relaxed. As we shall find it is the
relaxation of this condition which, in our approach, facilitates the
introduction of matter fields into the $4D$\ spacetime from higher
dimensions.

With the form of the above line element one finds that the only surviving
components of the $5D$ Ricci tensor, $R_{ab}$, are: 
\begin{eqnarray}
R_{00} &=&e^{\nu -\mu }\left[ \frac{\nu ^{\prime \prime }}{2}+\frac{\nu
^{\prime 2}}{4}-\frac{\nu ^{\prime }\mu ^{\prime }}{4}+\frac{\nu ^{\prime
}\psi ^{\prime }}{4}+\nu ^{\prime }\frac{R^{\prime }}{R}\right] +\varepsilon
e^{\nu -\psi }\left[ -\frac{\overset{\diamond \diamond }{\nu }}{2}-\frac{%
\overset{\diamond }{\nu }^{2}}{4}-\frac{\overset{\diamond }{\nu }\overset{%
\diamond }{\mu }}{4}+\frac{\overset{\diamond }{\nu }\overset{\diamond }{\psi 
}}{4}-\overset{\diamond }{\nu }\frac{\overset{\diamond }{R}}{R}\right] ; 
\notag \\
R_{11} &=&-\frac{\nu ^{\prime \prime }}{2}-\frac{\psi ^{\prime \prime }}{2}-%
\frac{\nu ^{\prime 2}}{4}-\frac{\psi ^{\prime 2}}{4}+\frac{\nu ^{\prime }\mu
^{\prime }}{4}+\frac{\nu ^{\prime }\psi ^{\prime }}{4}+\mu ^{\prime }\frac{%
R^{\prime }}{R}-\varepsilon e^{\mu -\psi }\left[ -\frac{\overset{\diamond
\diamond }{\mu }}{2}-\frac{\overset{\diamond }{\mu }^{2}}{4}+\frac{\overset{%
\diamond }{\nu }\overset{\diamond }{\mu }}{4}+\frac{\overset{\diamond }{\nu }%
\overset{\diamond }{\psi }}{4}-\overset{\diamond }{\nu }\frac{\overset{%
\diamond }{R}}{R}\right] ;  \notag \\
R_{22} &=&1-R^{2}e^{-\mu }\left[ \left( \frac{R^{\prime }}{R}\right) ^{2}+%
\frac{R^{\prime \prime }}{R}+\frac{R^{\prime }}{2R}(\nu ^{\prime }-\mu
^{\prime }+\psi ^{\prime })\right] +\varepsilon R^{2}e^{-\mu }\left[ \left( 
\frac{\overset{\diamond }{R}}{R}\right) ^{2}+\frac{\overset{\diamond
\diamond }{R}}{R}+\frac{\overset{\diamond }{R}}{2R}(\overset{\diamond }{\nu }%
-\overset{\diamond }{\mu }+\overset{\diamond }{\psi })\right] ;  \notag \\
R_{33} &=&\sin ^{2}\theta \left( R_{22}\right) ;  \TCItag{5} \\
R_{44} &=&-\frac{\overset{\diamond \diamond }{\nu }}{2}-\frac{\overset{%
\diamond }{\nu }^{2}}{4}-\frac{\overset{\diamond \diamond }{\mu }}{4}-\frac{%
\overset{\diamond }{\mu }}{4}^{2}+\frac{\overset{\diamond }{\nu }\overset{%
\diamond }{\psi }}{4}+\frac{\overset{\diamond }{\nu }\overset{\diamond }{%
\psi }}{4}-\overset{\diamond }{\psi }\frac{\overset{\diamond }{R}}{R}-\frac{%
\overset{\diamond \diamond }{R}}{2R}+\varepsilon e^{\psi -\mu }\left[ \frac{%
\psi ^{\prime \prime }}{2}+\frac{\psi ^{\prime 2}}{4}+\frac{\nu ^{\prime
}\psi ^{\prime }}{4}+\frac{\mu ^{\prime }\psi ^{\prime }}{4}+\psi ^{\prime }%
\frac{R^{\prime }}{R}\right] ;  \notag \\
R_{14} &=&-\frac{\overset{\diamond }{\nu }^{\prime }}{2}-\frac{\overset{%
\diamond }{\nu \nu }^{\prime }}{4}+\frac{\overset{\diamond }{\mu }\nu
^{\prime }}{4}+\frac{\overset{\diamond }{\nu }\psi ^{\prime }}{4}+\overset{%
\diamond }{\mu }\frac{R^{\prime }}{R}+\psi ^{\prime }\frac{\overset{\diamond 
}{R}}{R}-\frac{\overset{\diamond }{2R}^{\prime }}{R}.  \notag
\end{eqnarray}
Here the over head diamond `$\diamond $' means differentiation with respect
to the fifth coordinate while the prime `$\prime $', as usual, denotes
differentiation with respect to the radial coordinate.

Now, conditions (i) and (ii) suggest that, as an ansatz, we take a $5D$ line
element of the form 
\begin{equation}
dS^{2}=\left( 1-\frac{h}{r}\right) dt^{2}-\left( 1-\frac{h}{r}\right)
^{-1}dr^{2}-r^{2}d\Omega ^{2}-\varepsilon \phi dh^{2},  \tag{6}
\end{equation}
where $\phi =\phi \left( r,h\right) $ is a lapse function and $\varepsilon
=\pm 1$, as mentioned before, identifies the $h$ coordinate as either
spacelike or timelike. In this particular work we assume a coordinate system
in which $\phi $ is scaled to unity.

Using Eqs. (5) and (6) we find the only surviving components of the 5D Ricci
tensor $R_{ab}$ are 
\begin{eqnarray}
R_{00} &=&\frac{-1}{2\varepsilon r^{2}\left( 1-\frac{h}{r}\right) };  \notag
\\
R_{11} &=&\frac{-1}{2\varepsilon r^{2}\left( 1-\frac{h}{r}\right) ^{3}}; 
\notag \\
R_{14} &=&\frac{-1}{2r^{2}\left( 1-\frac{h}{r}\right) };  \TCItag{7} \\
R_{44} &=&\frac{1}{2r^{2}\left( 1-\frac{h}{r}\right) ^{2}}.  \notag
\end{eqnarray}
Further, the $5D$ Ricci scalar is given as 
\begin{equation}
R=R_{a}^{a}=\frac{1}{2r^{2}\varepsilon \left( 1-\frac{h}{r}\right) ^{2}} 
\tag{8}
\end{equation}
Using Eqs. (7) and (8) above to build the $5D$ Einstein tensor $%
G_{ab}=R_{ab}-\frac{1}{2}Rg_{ab}$ we find its surviving components to be

\begin{eqnarray}
G_{00} &=&\frac{-3}{4\varepsilon r^{2}\left( 1-\frac{h}{r}\right) };  \notag
\\
G_{11} &=&\frac{-1}{4\varepsilon r^{2}\left( 1-\frac{h}{r}\right) ^{3}}; 
\notag \\
G_{14} &=&\frac{-1}{2r^{2}\left( 1-\frac{h}{r}\right) };  \notag \\
G_{22} &=&\frac{r^{2}}{4\varepsilon r^{2}\left( 1-\frac{h}{r}\right) ^{2}}; 
\TCItag{9} \\
G_{33} &=&\sin ^{2}\theta G_{22};  \notag \\
G_{44} &=&\frac{1}{4r^{2}\left( 1-\frac{h}{r}\right) }  \notag
\end{eqnarray}

\section{The 4+1 splitting and the induced fields}

\smallskip In order to isolate physical information from the above full 5D
results it is worthwhile comparing such results with those from a $4+1$
splitting of the Kaluza-Klein theory. This comparison will manifest features
in the preceding results which suggest the existence of two distinct fields
induced on the 4D hyperface. To this end we start with an overview of a
Kaluza-Klein theory with the \textit{cylinder condition} relaxed, i.e. $%
\partial _{\left( a=4\right) }\left[ g_{\mu \nu }\right] \neq 0$. In such an
approach one can, in general, institute a $4+1$ split of the $5D$ metric $%
dS^{2}=g_{ab}dx^{a}dx^{b}$. This foliation leaves an induced 4D metric, $%
g_{\mu \nu }\left( x^{a}\right) ,\ (\mu ,\nu =0,..,3)$ which can be related
to the $5D$ metric by$^{11}$ 
\begin{equation}
g_{ab}=\left( 
\begin{array}{cc}
g_{\mu \nu } & N_{\mu } \\ 
N_{\nu } & \varepsilon \phi ^{2}+N_{\alpha }N^{\alpha }
\end{array}
\right) ,  \tag{10}
\end{equation}
where $N^{\mu }\left( x^{a}\right) $ is a shift vector (historically$^{3}$
associated with the electromagnetic fields).

The $5D$ theory can then be related to its $4D$ sector through a system of
15 equations which can be divided into three groups$^{12}$. In this
treatment we only discuss gravitational fields and as a result we are only
interested in the one system of 10 equations for the 4D Ricci tensor, 
\begin{equation}
R_{\mu \nu }=\frac{1}{\phi }\left[ D_{\mu }D_{\nu }\phi +\varepsilon \left( 
\mathit{L}_{N}K_{\mu \nu }-\partial _{h}K_{\mu \nu }\right) +\varepsilon
\phi \left( KK_{\mu \nu }-2K_{\mu \lambda }K_{\,\,\nu }^{\lambda }\right) %
\right] ,  \tag{11}
\end{equation}
where $K_{\mu \nu }=\frac{1}{2\phi }\left[ D_{\mu }N_{\nu }+D_{\nu }N_{\mu
}-\partial _{h}g_{\mu \nu }\right] $ is the extrinsic curvature tensor
induced on the 4D spacetime, $\mathit{L}_{N}K_{\mu \nu }$ is its Lie
derivative and $D_{\mu }$ is a covariant derivative operator. For the same
reason we gauge the $N^{\alpha }$ out and choose a coordinate system with $%
\phi =1$. With this the only surviving contribution to the extrinsic
curvature term, 
\begin{equation}
K_{\mu \nu }=-\frac{1}{2}\partial _{h}g_{\mu \nu },  \tag{12}
\end{equation}
results from the relaxation of the \textit{cylinder condition}.

One can now apply this approach to our ansatz (Eq. (6)). Here, the $4+1$
foliation of the full $5D$ spacetime gives a family of $4D$ hyperfaces with
an induced metric $g_{\mu \nu }$ whose line element is 
\begin{equation}
ds^{2}=\left( 1-\frac{h}{r}\right) dt^{2}-\left( 1-\frac{h}{r}\right)
^{-1}dr^{2}-r^{2}d\Omega ^{2},  \tag{13}
\end{equation}
and which we also take to be the physical metric$^{\ast }$\footnote{$^{\ast
} $In general, the induced metric $\grave{g}_{\mu \nu }$ and the physical
metric $g_{\mu \nu }$ can be related by $\grave{g}_{\mu \nu }\left(
r,h\right) =\Omega \left( r,h\right) g_{\mu \nu }\left( r,h\right) $, where $%
\Omega \left( r,h\right) $ is a warp factor.}. Then from Eqs. (12) and (13)
the explicit extrinsic curvature of the induced $4D$ spacetime is 
\begin{equation}
K_{\mu \nu }=-\frac{\varepsilon }{2}\partial _{h}\left[ g_{\mu \nu }\left(
x^{\alpha },h\right) \right] =\frac{-\varepsilon }{2r}diag.\left[ 
\begin{array}{cccc}
1, & \frac{1}{\left( 1-\frac{h}{r}\right) }, & 0, & 0
\end{array}
\right] ,  \tag{14}
\end{equation}
and, clearly, the associated curvature scala vanishes, 
\begin{equation}
K=g^{\mu \nu }K_{\mu \nu }=0.  \tag{15}
\end{equation}
Using Eqs. (14) and (15) in (11) one finds the components of the 4D Ricci
tensor to be 
\begin{equation}
R_{\mu \nu }=\frac{-\varepsilon }{2r^{2}\left( 1-\frac{h}{r}\right) ^{2}}%
diag.\left[ 
\begin{array}{cccc}
\left( 1-\frac{h}{r}\right) , & \frac{1}{\left( 1-\frac{h}{r}\right) }, & 0,
& 0
\end{array}
\right] .  \tag{16}
\end{equation}
The resulting Ricci scala is, manifestly traceless, 
\begin{equation}
R=-\left[ g^{\mu \nu }\left( \partial _{h}K_{\mu \nu }\right) +2g^{\alpha
\nu }K_{\alpha }^{\mu }K_{\mu \nu }\right] =0.  \tag{17}
\end{equation}

Consequently, Eqs.(16) and (17) recover the information originally contained
in the $4D$ sector of (5D Ricci tensor, Eq. (7)). Now, using Eqs. (16), (17)
and (13) we can construct a 4D Einstein tensor $G_{\mu \nu }$, and through
the Bianchi identities, the associated 4D field equations $G_{\mu \nu
}=-\tau _{\mu \nu }$ give the matter fields as 
\begin{equation}
\tau _{\mu \nu }=\frac{1}{2\varepsilon r^{2}\left( 1-\frac{h}{r}\right) ^{2}}%
diag\left[ \left( 1-\frac{h}{r}\right) ,\frac{1}{\left( 1-\frac{h}{r}\right) 
},0,0\right] .  \tag{18}
\end{equation}

It is apparent that the result in Eq. (18) is different from that expressed
by the $4D$ sector of the $5D$ Einstein tensor in Eq. (9) and which we write
as $\tilde{G}_{\mu \nu }$. The difference arises because the $R_{44}$
component in the $5D$ theory contributes to the non-vanishing of the $5D$
trace in Eq. (8). This then projects onto the $4D$ sector a cosmological
substrate fluid $\theta _{\mu \nu }$. The result is that the total matter
fields $T_{\mu \nu }$ in the 4D sector of $G_{ab}$ (Eq. (9)) can now be
represented as a sum of two fluids, 
\begin{equation}
T_{\mu \nu }=\tau _{\mu \nu }+\theta _{\mu \nu },  \tag{19}
\end{equation}
with $\tau _{\mu \nu }$ given by Eq. (18) and with $\theta _{\mu \nu }=$ $%
\theta g_{\mu \nu }$, where $\theta $ is obtainable from 5D Ricci scalar Eq.
(8) as 
\begin{equation}
\theta =\frac{1}{2}g^{44}R_{44}=\frac{1}{4r^{2}\varepsilon \left( 1-\frac{h}{%
r}\right) ^{2}}.  \tag{20}
\end{equation}

\section{Properties of the induced fields}

We now take a brief look at some features of the fields $\theta _{\mu \nu }$
and $\tau _{\mu \nu }$ obtained in the preceding section and comment on the
potential implications. On the 4D hyperface both these fields would fill the
region of the black hole spacetime usually referred to as the external
Schwarzschild solution.

As we have noted above the $\theta _{\mu \nu }$ field originates from the
non-vanishing contribution of $R_{4}^{4}$ to the 5D trace. It is a substrate
field with a cosmological character. The field represents a perfect fluid as
can be inferred by writing $\theta _{\mu \nu }$ in the perfect fluid form $%
\theta ^{\mu \nu }=\left( \rho _{\theta }+p_{\theta }\right) u^{\mu }u^{\nu
}-p_{\theta }g^{\mu \nu }$, with $u^{0}=\left( -g_{00}\right) ^{-\frac{1}{2}%
}=\left( 1-\frac{h}{r}\right) ^{-\frac{1}{2}}$ and $u^{i=1,2,3}=0$. Then one
easily verifies that the field behaves like a cosmological fluid with a
negative pressure and an equation of state 
\begin{equation}
p_{\theta }=-\rho _{\theta }.  \tag{21}
\end{equation}

On the other hand, the $\tau _{\mu \nu }$ field comes from the induced 4D
Ricci tensor. This field is traceless and represents an anisotropic fluid
with the only surviving pressure term being in the radial component.
Further, there is no energy transport, as indicated in Eq. (18) by the
absence of off-diagonal (momentum) terms. All these features can be made
apparent by writing the field in a quasi-perfect fluid form 
\begin{equation}
\tau _{\;}^{\mu \nu }=\frac{1}{\varepsilon r^{2}}\left( \frac{1}{1-\frac{h}{r%
}}\right) ^{2}u^{\mu }u^{\nu }-\frac{1}{2\varepsilon r^{2}}\left( \frac{1}{1-%
\frac{2h}{r}}\right) ^{2}\left[ \delta _{0}^{\mu }\delta _{0}^{\nu }+\delta
_{1}^{\mu }\delta _{1}^{\nu }\right] ,  \tag{22}
\end{equation}
with $u^{0}=\left( -g_{00}\right) ^{-\frac{1}{2}}=\left( 1-\frac{h}{r}%
\right) ^{-\frac{1}{2}}$ and $u^{i=1,2,3}=0$. One can infer from Eq. (22)
that the radial pressure satisfies the equation of state, 
\begin{equation}
p_{\tau }=\rho _{\tau },  \tag{23}
\end{equation}
where the fluid density $\rho _{\tau }$ and the pressure $p_{\tau }$ are
functions of the radial coordinate and are, respectively, given by 
\begin{equation}
\rho _{\tau }=\frac{1}{2\varepsilon \left( r-h\right) ^{2}},  \tag{24}
\end{equation}
and 
\begin{equation}
p_{\tau }=\frac{1}{2\varepsilon \left( r-h\right) ^{2}}.  \tag{25}
\end{equation}

Both the energy density (Eq. 24) and pressure (Eq. 25) are positive definite
with the result that the fluid is a `normal' field, obeying all the standard
energy conditions$^{13}$. The pressure $p_{\tau }$ supports the fluid,
holding it in hydrostatic equilibrium by providing an outward force $F\left(
r\right) $, at any point in the region outside the horizon. This force is
given by 
\begin{equation}
F\left( r\right) =-\frac{1}{\rho _{\tau }}\frac{\partial p_{\tau }}{\partial
r}=\frac{2}{r-h},  \tag{26}
\end{equation}
and is, clearly, due to a non-vanishing pressure gradient $\frac{\partial
p_{\tau }}{\partial r}$, in the radial direction. The force, $F\left(
r\right) $, whose origins are in the higher dimension, is nevertheless
gravitational in nature as can be verified by setting the gravitational
coupling constant $G$ to zero.

\section{\ \ \ Conclusion}

In this paper we have explored the possibility that spacetime horizons may
be manifestations of higher dimensions and that such objects may be
generators of energy fields in our 4D spacetime. Treating the length
associated with the horizon size as an independent fifth coordinate in an
uncompactified 5D theory we have constructed a set of solutions, one in 5D
gravity and its 4D counterpart in Einstein gravity sector for a black hole
with a horizon (length) size $h=\frac{2GM}{c^{2}}$.

The 4D solution shows the space outside a black hole to be filled with two
distinct fluids. First, we find a substrate field projected on the 4D from
the fifth dimension by the non-vanishing 5D trace element, $R_{4}^{4}$. The
field has a cosmological character, behaving as a perfect fluid with an
equation of state, $p_{\tau }=-\rho _{\tau }$, similar to that of a
cosmological constant. It differs from the cosmological constant only in its
dependence on the radial coordinate. We interpret the influence of the
negative pressure to imply that for the hole, the external 4D spacetime
appears to expand away. Conversely, for an observer in the external 4D
spacetime, the black hole would appear to contract away with the radial
coordinate.

\smallskip The second field, on the other hand, is matter-like in that it
originates from the non-vanishing components of the 4D sector of the Ricci
tensor. Because the two angular components of this tensor vanish, the
resulting fluid is anisotropic, having only a radial pressure component.
Both the radial pressure and the density are functions of the radial
coordinate and obey an equation of state of the form $p_{\tau }\left(
r\right) =\rho _{\tau }\left( r\right) $. The pressure support holds the
fluid in hydrostatic equilibrium with a radial force $F\left( r\right) =-%
\frac{1}{\rho _{\tau }}\frac{\partial p_{\tau }}{\partial r}$ at each point.
This force does not, however, constitute a ``fifth force'' as can be seen
from its coupling constant. The force is gravitational and vanishes on
setting Newton's gravitational constant $G$ to zero.

One notes that our 4D solution is a generalization of the Schwarzschild
solution, but only for a black hole. It cannot, for example, describe the
outside of a regular star, whether radiating$^{14}$ or not$^{8}$. This is
because the mass of a regular star occupies a physical 4-volume and
therefore cannot be treated as an independent coordinate. Taken as they are,
these results would suggest possible non-local consequences on the geometry
of spacetime resulting from the collapse of matter to form a black hole
horizon (and the implied eventual future-directed singularity). The physical
implications, then, would be that the global geometry of the spacetime
outside a black hole differs from that of an uncollapsed star of the same
mass.

In closing we mention some new features and implications that the general
treatment of mass as a fifth coordinate makes manifest. One of the
improvements that Einstein made to the theory of gravitation (see for
example MTW)$^{15}$ was the realization that Newton's concept of gravity as
a force could be replaced by spacetime curvature. This approach did, among
other things, solve the problem of action at a distance in Newton's theory.
In our present approach, the view of gravity as inducing curvature on the 4D
manifold does not change. The only inference that our approach makes is that
spacetime curvature and ordinary matter/radiation fields in the 4D spacetime
are manifestations of a higher dimension, resulting from the mixing of the
fifth coordinate with the 4D spacetime coordinates. There are two other
interesting implications of this approach. The first relates to the expected
effects of the fifth dimension on the geometry of the universe. Consistent
with the foregoing arguments one expects that the cosmological effect of the
fifth dimension will be a cosmological field with a non-vanishing energy
density induced on the 4D sector of the 5D manifold. Consequently, the
global metric defined on the 4D universe may not be asymptotically flat.
This point, which will be taken up again in a future discussion, does seem
to naturally justify the cosmological constant and possibly predict
inflation. Lastly, taking mass as a dynamical coordinate is a stepping stone
(as our solutions in Eqs. (6) and (13) indicate) towards a scale invariant
theory of nature.

\begin{acknowledgement}
This work was made possible by funds from the University of Michigan.
\end{acknowledgement}

\end{document}